\documentstyle[twocolumn,aps,epsf]{revtex}

\newcommand{\be}[1]{\begin{equation} \label{(#1)}}
\newcommand{\ee}{\end{equation}}
\newcommand{\ba}[1]{\begin{eqnarray} \label{(#1)}}
\newcommand{\ea}{\end{eqnarray}}
\newcommand{\nn}{\nonumber}
\newcommand{\rf}[1]{(\ref{(#1)})}

\begin{document}
\draft
\title{Test of Special Relativity and Equivalence Principle\\ 
from Neutrinoless 
Double Beta Decay}
\author{H.V. Klapdor--Kleingrothaus$^1$, H. P\"as$^1$, U. Sarkar$^{2,3}$}
\address{$^1$ Max--Planck--Institut f\"ur Kernphysik
P.O. Box 103980,
D--69029 Heidelberg,
Germany}
\address{$^2$ DESY, Notkestrasse 85, D--22607 Hamburg, Germany}
\address{$^3$ Physical Research Laboratory, Ahmedabad, 380 009, India}
\date{\today}
\maketitle
\begin{abstract}
We generalize the formalism for testing Lorentz invariance and the
equivalence principle in the neutrino sector. While neutrino oscillation
bounds constrain the region of large mixing of the the weak and
gravitational eigenstates, we obtain
new constraints on violations of Lorentz invariance and the equivalence 
principle from neutrinoless double beta decay. 
These bounds apply even in the case of no mixing 
and thus probe a totally unconstrained region in the parameter 
space. 
\end{abstract}

% body of paper here
Special relativity 
and the equivalence principle can be considered as the most 
basic foundations of the theory of gravity. 
Many experiments already have tested these principles to a very high 
level of
accuracy \cite{rel} for ordinary matter - generally for 
quarks and leptons of the first
generation. These precision tests of 
local Lorentz invariance -- violation of the equivalence 
principle should produce a similar effect \cite{will} -- probe for any 
dependence of the (non--gravitational) laws of physics on a laboratory's 
position, orientation or velocity relative to some preferred frame of
reference, such as the frame in which the cosmic microwave background is 
isotropic.  

A typical feature of the violation of local Lorentz invariance (VLI)
is that different species of matter have a characteristical 
maximum attainable speed.
This can be tested in various sectors of the standard model
through vacuum Cerenkov radiation \cite{gasp}, photon decay \cite{cole},
neutrino oscillations \cite{glash,nu1,nu2,hal,nu3} and $K-$physics
\cite{hambye,vepk}. In this article we extend
these arguments to derive new constraints from neutrinoless double
beta decay. 

The equivalence principle implies that spacetime is described by
unique operational geometry and hence universality of the gravitational 
coupling for all species of matter. In the recent years there
have been attempts to constrain a possible amount of 
violation of the equivalence principle (VEP) in the neutrino sector
from neutrino oscillation experiments \cite{nu1,nu2,hal,nu3}.
However, these bounds don't apply when the gravitational and the
weak eigenstates have small mixing. In this article we
present a generalized formalism of the neutrino sector to test the VEP
and point out that neutrinoless double beta decay also constrains the 
VEP. VEP implies different neutrino species to suffer from  
different gravitational potentials while propagating through the 
nucleus and hence the effect of different eigenvalues doesn't cancel
for the same effective momentum. Earlier results on neutrino oscillations 
come out as special case from our present formalism.
The main result is that neutrinoless double beta decay can constrain
the amount of VEP even when the mixing angle is zero, {\it i.e.},
when only the weak equivalence principle is violated, for which 
there does not exist any bound at present.

We shall first present our formalism for VLI and then for VEP.
For sake of clarity we formulate the problem for a
two generation scenario involving $\nu_e$ and $\nu_x$ with 
$x=\mu,\tau,s$.
Neutrinos of different species may have different maximum attainable 
velocities if there is violation of local Lorentz invariance
(VLI) and hence violation of special relativity \cite{cole}.
We first assume that the weak eigenstates cannot be diagonalized 
simultaneously with the velocity eigenstates and the neutrinos are
relativistic point particles. 
The effective Hamiltonian in the weak basis $[\nu_e ~~ \nu_x ]$ is
\begin{equation}
H = U_{m} H_{m} U_{m}^{-1} + U_v H_v U_v^{-1} \label{h}.
\end{equation}
In absence of VLI the neutrino mass matrix
in the mass basis $[\nu_1 ~~ \nu_2 ]$ is given by
\be{hsew}
H_{m} = \frac{(M_{m})^2}{2 p} = \frac{1}{2 p} {\pmatrix{
m_1 & 0 \cr 0 & m_2 }}^2 \label{hsew}
\end{equation}
and the VLI part of the hamiltonian as
\begin{equation}
H_v = \pmatrix{ v_1 & 0 \cr 0 & v_2} p  , 
\ee
to leading order in $\bar{m}^2/p^2$. Here p denotes the momentum and
$\bar{m}$ the average mass, and for any quantity $X$ we define $\delta X
\equiv(X_1-X_2)$, $\bar{X} = (X_1+X_2)/2$. 

In the absence of VLI, {\it i.e.}, when the special threory of
relativity is valid, $v_i = 1$, and $H_v$ simply becomes the momentum
of the neutrinos. Here we are interested in a single neutrino
beam (for neutrino oscillation experiments) or a single virtual neutrino
propagating inside the nucleus with a particular momentum. For this 
reason we assume the momenta of both the neutrinos are $p$. Then
$v_i $ corresponds to the maximum attainable speed of the corresponding
momentum eigenstates. Hence $v_1 - v_2 = \delta v$ is a measure of 
VLI in the neutrino sector. As typical or ``standard'' maximum attainable 
speed $\frac{v_1+v_2}{2}=1$ is assumed. All previous bounds on this quantity
$\delta v$ in the neutrino sector were derived from neutrino oscillation
experiments and for that reason these bounds are valid only for 
large gravitational mixing. As we shall point out, 
neutrinoless double beta decay can constrain
$\delta v$ even when the mixing angle vanishes. 

We shall not consider any $CP$ violation, and hence $H_{m}$ and
$H_v$ are real symmetric matrices and $U_{m}$ and $U_v$ are orthogonal
matrices $U^{-1}= U^T$. They can be parametrized as 
$U_i = \pmatrix{\cos \theta_i & \sin \theta_i \cr -\sin \theta_i
& \cos \theta_i}$, where $\theta_i$ represents weak mixing angle $\theta_m$ 
or velocity mixing angle $\theta_v$. We can now write down the 
weak Hamiltonian $H_w$ in the basis $[\nu_e ~~ \nu_x]$, 
in which the charged lepton mass matrix is diagonal
and the charged current interaction is also diagonal, as
$$
H = p I +  \frac{1}{2 p} { \pmatrix{ M_{+} & M_{12} \cr
M_{12} & M_{-} }}^2.
$$
Here $I$ is the identity matrix and
\ba{obs}
M_{\pm} &=& \bar{m} \pm \frac{cos2\theta_m}{2}
\delta m \nn \\
&&\pm \frac{p^2}{\bar{m}} \delta v \left( \frac
{\cos2\theta_v}{2} - {\delta m \over 4 \bar{m}} \cos 2 (\theta_m - \theta_v)
\right) \nonumber \\
M_{12} &=& -{\sin2\theta_m \over 2} \delta m\nn \\
&&- \frac{p^2}{\bar{m}} \delta v \left( {\sin2\theta_v \over 2} 
+ {\delta m \over 4 \bar{m}} \sin 2 (\theta_m - \theta_v) \right) .
\ea

In the case of exact Lorentz invariance, 
we usually write the mass matrix in the weak basis 
as $\pmatrix{ m_{ee} & m_{e \mu} \cr m_{e \mu} & m_{\mu \mu}}$. 
In the mass mechanism of neutrinoless double beta decay, 
the decay rate
\be{t12}
[T_{1/2}^{0\nu\beta\beta}]^{-1}=\frac{\langle m \rangle^2}{m_e^2} 
G_{01} |ME|^2,
\ee
is proportional to the effective neutrino mass 
$\langle m \rangle=m_{ee}=M_+$.
Here $ME$ denotes the nuclear matrix element $ME=M_F-M_{GT}$, $G_{01}$
corresponds to the phase space factor defined in \cite{doi} and $m_e$ is the 
electron mass.
The double beta observable can be written as
\ba{1}
<m> &=& \sum_i U_{e i}^2 m_i = m_1 \cos^2 \theta_w + m_2 \sin^2 \theta_w\nn \\
&=& \bar{m} + {1 \over 2} \delta m \cos 2 \theta_w.
\ea
If $m_{ee}= 0$, the two physical eigenstates with
eigenvalues $m_1$ and $m_2$ will contribute to the neutrinoless 
double beta decay by an amount $U_{e1}^2 m_1$ and $U_{e2}^2 m_2$,
respectively, which cancels each other. However, if these two physical 
states have different maximum attainable speed, corresponding to VLI,
this cancellation 
will not be exact for the same cut--off effective momentum in the neutrino 
propagator. As a result, even when 
$m_{ee}= 0$, we can have neutrinoless double beta decay, which is
proportional to the amount of VLI and the double beta observable is
given by $M_+$ in \rf{obs}. From \rf{obs} it can easily be seen that in the 
region of maximal mixing
$\cos 2 \theta_v = 0$, the double beta decay rate vanishes. 
Thus neutrinoless double beta decay doesn't constrain
the amount of VLI for maximal mixing. However, when the mixing approaches zero,
the most stringent bound from neutrinoless double 
beta decay is obtained. 
In this case $\delta{v}/2$ can be understood as derivation from the standard
maximum attainable speed $\bar{v}$.
As it is obvious, when there is no mixing the
neutrino oscillation experiments cannot give any bound on the amount of 
VLI, since in absence of mixing only VLI cannot allow neutrino 
oscillations. 

To give a bound on VLI in the small mixing region 
(including $\theta_v=\theta_m=0$)
we assume conservatively
$\langle m \rangle \simeq 0$.
We also assume 
$\delta{m} \leq \bar{m}$, and thus $\frac{\delta m}{4 \bar{m}}$ may be 
neglected. 
Due to the $p^2$ enhancement the nuclear matrix elements of the 
mass mechanism have to be replaced by $\frac{m_p}{R}\cdot 
(M_F^{'}-M_{GT}^{'})$ with the nuclear radius $R$ and the proton mass $m_p$, 
which have been 
calculated in \cite{mat}.
Inserting the recent half life limit obtained from the Heidelberg--Moscow 
experiment \cite{double}, $T_{1/2}^{0\nu\beta\beta}>1.2\cdot 10^{25} y$,  
a bound on the amount of VLI as a function of the average neutrino mass
$\bar{m}$ can be given. The most reliable assumption for $\bar{m}$ is obtained 
from the cosmological bound $\sum_i m_i < 40$ eV \cite{raf}, i.e., 
$\bar{m}<13$ eV for three generations, implying a bound of
$$\delta v < 4 \times 10^{-16}~~~~ {\rm for}~~~ \theta_v=\theta_m =0.$$ 
However, combining the present experimental constraints from atmospheric and 
solar neutrino data as well as from tritium beta decay in a three neutrino 
framework and assuming a
typical hierarchical mass pattern spectrum 
$m_3\gg m_{1,2}$ or $m_3\simeq m_2\gg m_1$ implies 
$\bar{m}<\sim 0.08$ eV
\cite{bar}
and improves the bound to $\delta v< 2 \cdot 10^{-18}$ for
$\theta_v=\theta_m =0$.
  
In figure 1 the bound implied by double beta decay
is presented for the entire range of $sin^2 2 \theta_v$ and compared with
bounds obtained from neutrino oscillation experiments in \cite{hal}. 
It should be stressed also that the GENIUS proposal of the Heidelberg group
\cite{gen} could improve these bounds  
by about 1--2 orders of magnitude. 
 
For comparison, in the following the amount of VLI in neutrino oscillation 
experiments is calculated in this formalism.
 In the basis of the
physical states $\nu_a$ and $\nu_b$, the Hamiltonian becomes
\ba{2}
H &=& \pmatrix{p + \frac{m_a^2}{2 p} & 0 \cr 0 & p + \frac{m_b^2}{2 p} }\nn \\
&=& \pmatrix{\bar{E} & 0 \cr 0 &
 \bar{E}} + {1 \over 2}
\pmatrix{ \Delta E & 0 \cr 0 & - \Delta E } \label{he},
\ea
where $\bar{E} =  ( p + \frac{\bar{m}^2}{2 p} )$
and
\ba{mls}
&&\frac{p}{\bar{m}} \Delta E = m_a - m_b  \nn \\
&&= \left[
( \delta m )^2 + {\left( \delta v {p^2 \over \bar{m}}
\right)}^2 + 2 \delta m \delta v \frac{p^2}{\bar{m}}
\cos(2(\theta_w - \theta_v)) \right]^{1 / 2}.
\ea
The new mixing angle $\theta_{tot}$
is a function of $\theta_m$ and $\theta_v$ and
the oscillation probability is now given by
\begin{equation}
P(\nu_e \to \nu_x) = \sin^2 2 \theta_{tot} \sin^2 {\pi L \over \lambda},
\end{equation}
where $\lambda = {\pi p \over \Delta m^2}$ and $\Delta m^2 =
m_a^2 - m_b^2$. In the limit of vanishing neutrino masses, 
neutrino oscillations are implied only by VLI. In this case the oscillation
probability becomes
$$ P(\nu_e \to \nu_x) = \sin^2 2 \theta_{v} 
\sin^2 { p L \delta v}, $$
which corresponds to the expression obtained earlier \cite{glash}. 
Here $p$ denotes
the total beam energy. From this expression it beomes clear that in the
case of no mixing, $\theta_v = 0$, neutrino oscillation experiments
don't constrain the size of VLI effects. 

In the following we present the formalism for violation of the equivalence 
principle (VEP). While in the final expression the amount of
VLI just will be replaced by VEP, the origin differs.  In a linearized 
theory the
gravitational part of the Lagrangian to first order in a 
weak gravitational field $g_{\mu\nu}=\eta_{\mu\nu}+    h_{\mu\nu}$
($h_{\mu\nu}= 2\frac{\phi}{c^2}  {\mbox diag}(1,1,1,1)$)
can be written as ${\cal L} = -\frac{1}{2}(1+g_i)h_{\mu\nu}T^{\mu\nu}$,
where  $T^{\mu\nu}$  is the  stress-energy  in the  gravitational
eigenbasis. In the presence of VEP the $g_i$ may differ. Assuming only 
violation of the weak
equivalence principle, the gravitational
interaction is diagonal but the couplings differ. In this case there 
does not exist any bound on the amount of VEP. We point out that this
region of the parameter region is most restrictively bounded by  
neutrinoless double beta decay.

The effective Hamiltonian in the weak basis again can be written as
\begin{equation}
H = p  I + U_m H_{m} U_m^{-1} + U_G H_G U_G^{-1}, \label{hha}
\end{equation}
with $H_m$ given in \rf{hsew} and
\ba{4}
H_G &=& \pmatrix{ G_1 & 0 \cr 0 & G_2} \nn \\
&=& \pmatrix{
- 2 (1 + g_1) \phi (p + \frac{\bar{m}^2}{2 p}) & 0 \cr 0 &
- 2 (1 + g_2) \phi (p + \frac{\bar{m}^2}{2 p})} \label{hga}
\ea
to first order in $\bar{m}^2/p^2$.
In formalisms where only violation of the weak equivalence principle is 
assumed, one
starts with $U_G$ proportional to $U_W$, in which case there does
not exist any bound from the neutrino oscillation experiments. 

The Hamiltonian in the weak basis is
\ba{5}
M_{\pm} &=& \bar{m} \pm {\cos 2 \theta_m \over 2} {\delta {m}} \nn \\
&&
\pm \frac{p}{\bar{m}} \delta G \left( \frac{\cos 2\theta_G}{2}
- {\delta m \over 4 \bar{m}} \cos (\theta_m - \theta_v) \right) \nonumber \\
M_{12} &=& -{\sin2\theta_m \over 2} \delta m \nn \\
&&
- \frac{p}{\bar{m}} \delta G \left( 
{\sin2\theta_G \over 2} + {\delta m \over 4 \bar{m}} \sin 
(\theta_m - \theta_v) \right)  .
\ea

Compared to VLI, the expressions for this case of VEP remain unchanged, 
except for replacing $\delta v$ by $\frac{1}{p}\delta G$.
Again, $\bar{g}=\frac{g_1+g_2}{2}$ can be considered as the standard 
gravitational coupling, for which the equivalence principle applies.

Thus the discussion of VLI can be directly translated to 
the VEP case and the bound from neutrinoless double beta
decay for $\theta_v=\theta_m =0$ is now given by
\ba{99}
\phi \delta g < 4 \times 10^{-16} {\rm GeV}~ ({\rm for~} \bar{m}<13 
{\rm eV})\nn \\
\phi \delta g < 2 \times 10^{-18} {\rm GeV}~ ({\rm for~} \bar{m}<0.08 
{\rm eV})
\ea
In this case, $\delta G = p \phi \delta g$, where $\phi$ is the 
background Newtonian gravitational potential on the surface of
the earth. A natural choice for $\phi$ would be the earth's 
gravitational potential $(\sim 10^{-9})$, but another well motivated 
choice could be 
the potential due to all forms of distant matter. Unlike the case of
VLI, the bound on the VEP will depend on what one chooses for the
Newtonian potential $\phi$. For this reason, here we only present
the combined bound on $\phi \delta g$.

In summary, we presented a general formalism for the study of both
VLI and VEP in the neutrino sector. We pointed out that 
neutrinoless double beta decay can constrain the amount of VLI or 
VEP. In particular, when the mixing of the gravitational 
eigenstates vanishes, the bounds from
neutrinoless double beta decay become most stringent, while this region is not 
constrained by any other experiments.

\vskip .1in {\bf Acknowledgement} 
One of us (US) would like to acknowledge the hospitality of the 
Theory Group, DESY, Hamburg, the Max-Planck-Institut f\"{u}r Kernphysik,
Heidelberg and a fellowship from the Alexander von Humboldt Foundation.

%\newpage

\newpage
\begin{figure}
\epsfysize=80mm
\vspace*{5mm}
\epsfbox{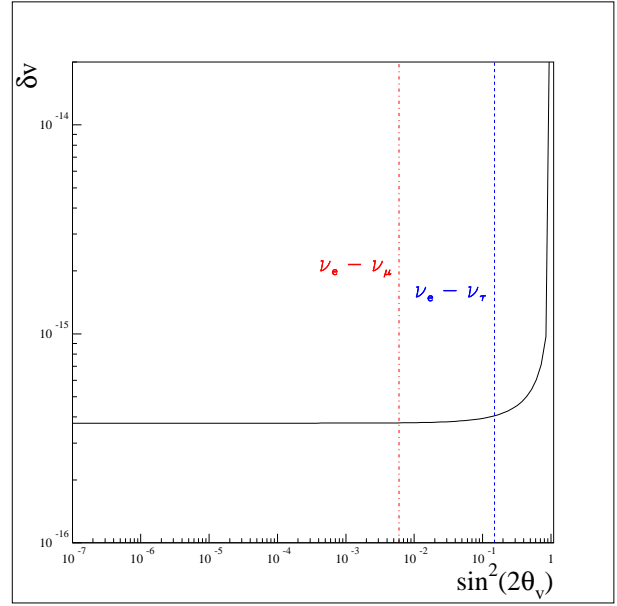}
\vspace*{5mm}
\caption{\it Double beta decay bound (solid line)
on violation of Lorentz invariance 
in the neutrino sector, excluding the region to the upper left. 
Shown is a double logarithmic plot 
in the $\delta v$--$\sin^2(2 \theta)$ parameter space. 
The bound becomes most stringent for the
small mixing region, which has not been constrained from any
other experiments. For comparison the bounds obtained from neutrino oscillation
experiments (from [10])
in the $\nu_{e} - \nu_{\tau}$ (dashed lines) and in the
$\nu_e - \nu_\mu$ (dashed-dotted lines) channel, excluding the region to the 
right, are shown. 
\label{1}}
\end{figure}

\end{document}